\documentstyle[11pt,amssymb]{article}
\newcommand{\ot}{\otimes}

\title{\large\bf{Separability of Multi-Partite Quantum States\footnotetext{*Corresponding author: jing@math.ncsu.edu}} ~ }
\author{{\small Xiaofen Huang$^{1}  $~~ Naihuan Jing$^{1, 2}$~*\footnotetext{}}\\
{\small {1. School of Sciences, South China University of
Technology, Guangzhou, Guangdong 510640, China}}\\{\small {2.
Department of Mathematics, North Carolina State University, Raleigh,
NC 27695, USA}}}
\date{}

\begin{document}
\maketitle
\begin{minipage}{123mm}

{\bf Abstract}~~{\small We give a direct tensor decomposition for
any density matrix into Hermitian operators. Based upon the
decomposition we study when the mixed states are separable and
generalize the separability indicators to multi-partite states and
show that a density operator is separable if and only if the
separable indicator is non-negative. We then derive two bounds for
the separable indicator in terms of the spectrum of the factor
operators in the tensor summands.}


\end{minipage}

\section{Introduction}

    In the last decade quantum entanglement has played a remarkable role in many applications and
    become one of the key resources in the rapidly expanding fields of quantum information and quantum
    computation, especially in quantum teleportation,
    quantum cryptography, quantum dense coding and parallel computation [1, 2, 3]. A quantum state or density
    matrix is separable
    (or not entangled) if it is a convex sum of tensor product of quantum states. In this case
    the separable quantum state can be
    prepared in several different locations.
    There are two aspects in the question regarding quantum entanglement:
    the first is to judge whether a general quantum state is entangled or not, and the second
    is to answer how much entanglement remained after some
    noisy quantum process. In the case of pure states, the Bell inequality provides a useful
    tool to tell separability from entanglement [4]. In [5, 6, 7, 8] separability problem was examined and
    important criteria were proposed from
    several viewpoints for the far more difficult case of
    mixed states including the PPT criterion and the range equality
    condition. In terms of measurement of entanglement other methods have been found, e.g. formation of entanglement [9, 10]
    and purification of formation [11, 12]. Recently further important and interesting works [13, 14, 15] have also
    been devoted solely to quantum entanglement and some
    criteria were proposed accordingly, in particular, [16]
    gives an operational and geometric approach to pairwise entanglement of
two and three-dimensional composite quantum systems.  Despite
these important developments the question of
    separability still remains unsolved and is notoriously famous for its
    difficulty.

Among the approaches to quantum separability it is highly needed to
have an operational method to decompose the quantum states as tensor
product. Such an idea was first studied in [17], where some
necessary constraints were found to ensure an optimal separable
approximation to a given density matrix, and then a numerical method
is proposed to locate the optimal separable state for two-partite
mixed states. In [18] a new algebraic mechanism was introduced to
study the separability question for two partite mixed states. The
idea was first to decompose the mixed density matrix as a summation
of tensor products of Hermitian operators, and then we rearrange the
sum to get the indicator. It was proved that the density matrix is
separable if and only if the indicator is non-negative. Thus the
indicator provides a new measurement for the separability.

In this article we will generalize this method to multi partite
density operators. We will give a new operational method to
decompose the density matrix as a summation of tensor products of
Hermitian operators. Our new method at the simplest case is the
fundamental fact that any $4\times 4$ Hermitian operator is a span
of composite Pauli spin matrices $\sigma_i\otimes \sigma_j$, where
$\sigma_0=\left(\begin{array}{cc} 1 & 0\\ 0 & 1\end{array}\right)$,
$\sigma_1=\left(\begin{array}{cc} 0 & 1\\ 1 & 0\end{array}\right)$,
$\sigma_2=\left(\begin{array}{cc} 0 & -i\\ i &
0\end{array}\right)$, and $\sigma_3=\left(\begin{array}{cc} 1 & 0\\
0 & -1\end{array}\right)$.

Once the decomposition into tensor product is known, the idea of
separability indicator [18] is generalized into multi-partite states
and we show that the mixed states are separable if and only if the
separability indicator is non-negative. In general it is hard to
compute the separability indicator. For this purpose we provide
several bounds, and hope that they will help in determination of the
separability.

\bigskip

\section{Basic notions}

  \vskip 0.1in
  Let $H_{1}$(resp. $H_{2}$)be an m(resp. n)-dimensional complex Hilbert space, with $|i\rangle, i=1,2 ... ,m$
  (resp. $|j\rangle, j=1, 2 ... ,n$) as an orthonormal basis. A bipartite mixed state is said to be separable
  if the density matrix can be written as
  $$\rho=\sum\limits_{i}p_{i}\rho_{i}^{1}\otimes\rho_{i}^{2}, \eqno(1) $$
  where $0<p_{i}\leq 1$, $\sum\limits_{i}p_i=1$,  $\rho_{i}^{1}$ and $\rho_{i}^{2}$ are density matrices
  on $H_{1}$ and $H_{2}$ respectively. It is a challenge problem to find such a decomposition or proving that it does
  not exist for a generic mixed state [5, 6, 7, 8].

     We first introduce some notations. For an $m\times m$ block matrix Z with each block $Z_{ij}$ of size
 $n\times n$, $i, j=1, 2 ... , m$. The realigned matrix $\tilde{Z}$ is defined by
  $$\tilde{Z}=[vec(Z_{11}), ... , vec(Z_{m1}), ..., vec(Z_{1m}), ..., vec(Z_{mm})]^{t}, \eqno (2) $$
  where for any $m\times n$ matrix T with entries $t_{ij}$, $vec(T)$ is defined to be
  $$vec(T)=[t_{11},..., t_{m1}, t_{12}, ..., t_{1n}, ..., t_{mn}]^{t}.$$

  Let $A=A^{R}+\sqrt{-1}A^{I}$ be a complex Hermitian matrix, where $A^{R}$ and $A^{I}$ are real
  and imaginary parts of $A$. Let $\sigma$ be the canonical map
from $A$ to a real matrix:
          $$\sigma:A \longmapsto \left(
                         \begin{array}{ll}
                          ~~A^{R}~~~~~A^{I}\\
                         -A^{I}~~~~~A^{R}\\
                          \end{array}
                          \right),  \eqno (3) $$
where $A^{R}$ and $A^{I}$ are the real and imaginary parts of A
respectively.
 \vskip 0.1in
   Let $Q_{s}$ be an $m^{2}\times \frac{m(m-1)}{2}$ matrix. If we arrange the row indices of $Q_{s}$
as $$\{11,21,31,...,m1,12,22,32,...,m2,...,mm\},$$
then all the
entries of $Q_{s}$ are zero except those at 21 and 12 (resp.31 and
13,...) which are 1 and -1 respectively in the first (resp.
second,...) column. In other words,
 $$Q_{s}=[\{e_{21},-e_{12}\};\{e_{31},-e_{13}\};...;\{e_{m,m-1},-e_{m-1,m}\}],$$
where $\{e_{21},-e_{12}\}$ is first column of $Q_s$, with $1$ and
$-1$ at the $21$ and $12$ rows respectively; while
$\{e_{31},-e_{13}\}$ is second column of $Q_s$, with $1$ and $-1$ at
the $31$ and $13$ rows respectively; and so on.

 \vskip 0.1in
 Let $Q_{a}$ be an $m^{2}\times\frac{m(m+1)}{2}$ matrix such that
 $$Q_{a}=[\{e_{11}\};\{e_{21},e_{12}\};\{e_{31},e_{13}\},...;\{e_{22}\};\{e_{32},e_{23}\},
 \{e_{42},e_{24}\};...;\{e_{m,m-1},e_{m-1,m}\},\{e_{mm}\}],$$
 where $\{e_{11}\}$ is the column vector with $1$ at the row $ii$ and zero elsewhere,
 and $\{e_{1j}, e_{1j}\}$ is the column vector with $1$ at the
 $ij$th and $ji$th rows and zero elsewhere.
\vskip 0.1in
 $Q_{1}$ can be expressed as
$$Q_{1}=\left( \begin{array}{ll}
                       \overline{Q_{s}}~~~~~~0~~~~~~0~~~~\overline{Q_{a}}\\
                          ~~0~~~~~\overline{Q_{a}}~~~~\overline{Q_{s}}~~~~0\\
                          \end{array}
                          \right),
                           $$
where $\overline{Q_{s}}$ and $\overline{Q_{a}}$ are obtained by
normalizing each column of $Q_{s}$ and $Q_{a}$.

By replacing the dimension m with n, we have $Q_{2}$.

   As an example we have for m=2
   $$ Q_{s}= \left( \begin{array}{llll}
                         0\\
                         1\\
                         -1\\
                         0\\
                          \end{array}
                          \right),
                           Q_{a}= \left( \begin{array}{llll}
                         1~~0~~0\\
                          0~~1~~0 \\
                          0~~1~~0\\
                           0~~0~~1\\
                          \end{array}
                          \right).  $$

\bigskip

\section{The tensor product decompositions
of Hermitian matrices}

\vskip 0.2in Let $A$ be a Hermitian matrix on Hilbert space
$H_1\otimes H_2$. In [18] we gave an operational method to decompose
$A$ as a tensor product of Hermitian matrices on $H_1$ and $H_2$. We
will give another method to decompose $A$ and then generalize to the
case of multi-tensor products.

Let's recall the decomposition method in [18]. We express the matrix
$A$ in terms of real and complex parts: $A=A^R+iA^I$ and realign
both $A^R$ and $A^I$ into $\tilde A^R$ and $\tilde A^I$ respectively
as in Eq. (2). Then we write
$$Q_{1}^{t}\left( \begin{array}{cc}
                        ~~\tilde{A^{R}}~~~\tilde{A^{I}} \\
                      {-\tilde{A^{I}}}~~\tilde{A^{R}}  \\
                          \end{array}
                          \right)Q_{2}=
                          \left( \begin{array}{ll}
                        \hat{A_{11}}~~\hat{A_{12}} \\
                      \hat{A_{21}}~~\hat{A_{22}}  \\
                          \end{array}
                          \right). \eqno (4)
                          $$

{\bf Proposition 1.} Let A be an $mn\times mn$
Hermitian matrix as rewritten in Eq. (4). Suppose the singular value
decomposition of $\hat{A_{22}}$ is
$\hat{A_{22}}=\sum_{i=1}^{r}\sqrt{\lambda_{i}}u_{i}v_{i}^{t}$, where
$r$ is the rank and $\lambda_{i}$(i=1,2...,r) are the non-zero
eigenvalues of $\hat{A_{22}}^{\dag}\hat{A_{22}}$, and $u_{i}$ (resp.
$v_{i}$) are the eigenvectors of the matrix
$\hat{A_{22}}\hat{A_{22}}^{\dag}$
(resp.$\hat{A_{22}}^{\dag}\hat{A_{22}}$). Set
$\hat{{\mathcal{B}}_{i} }=\sqrt{\lambda_{i}}u_{i}$,
$\check{\mathcal{C}}_{i}=-v_{i}.$ Then we can decompose $A$ as a
tensor product
$$A=\sum^r\limits_{i=1}B_{i}\otimes C_{i},$$
where the $m\times m$ Hermitian matrices
$B_i=b_i+\sqrt{-1}{\mathcal{B}}_{i}$
 and the $n\times n$ Hermitian matrices
$C_i=c_i+\sqrt{-1}{\mathcal{C}}_{i}$ are given by
$$\left( \begin{array}{c}
                        vec(b_{i}) \\
                      -vec({\mathcal{B}}_{i}) \\
                          \end{array}
                          \right)=Q_{1}\left( \begin{array}{c}
                        0 \\
                      -\hat{{\mathcal{B}}_{i} }\\
                          \end{array}
                          \right),~~
                          \left( \begin{array}{c}
                        vec(c_{i}) \\
                      vec({\mathcal{C}}_{i})  \\
                          \end{array}
                          \right)=Q_{2}\left( \begin{array}{c}
                        0 \\
                      \check{\mathcal{C}}_{i} \\
                          \end{array}
                          \right).   \eqno (5)$$

The above result gives a constructive or operative method to
decompose $A$ as a tensor product. The existence of tensor
decomposition has a simpler explanation. In fact, we know that the
set of $n\times n$ Hermitian matrices is a real vector space of
dimension $n^2$, thus the dimension of Hermitian matrices of size
$mn\times mn$ is exactly equal to the product of the dimension of
size $m\times m$ and that of size $m\times m$, hence the subspace of
tensor product of Hermitian matrices of size $n\times n$ and that of
size
 $m\times m$ must equal to the space of all Hermitian matrices
 of size $mn\times mn$, which
guarantees the existence.

We observe that in general the space of real symmetric
(antisymmetric) matrices can not be decomposed into tensor product
of symmetric (antisymmetric) matrices. In fact, the difference
between dimensions of the space of $mn\times mn$ symmetric matrices
and that of the tensor product of symmetric matrices of size
$m\times m$ and size $n\times n$ is
$$\left(\begin{array}{c}mn+1\\2\end{array}\right)-\left(\begin{array}{c}m+1\\2\end{array}\right)
\left(\begin{array}{c}n+1\\2\end{array}\right)=
\left(\begin{array}{c}m\\2\end{array}\right)\left(\begin{array}{c}n\\2\end{array}\right).$$
Similarly the difference between the dimensions of antisymmetric
operators over ${\mathbb C}^m\times  {\mathbb C}^n$ and that of the
tensor product of antisymmetric operators is
$$\left(\begin{array}{c}mn-1\\2\end{array}\right)-\left(\begin{array}{c}m-1\\2\end{array}\right)
\left(\begin{array}{c}n-1\\2\end{array}\right)=
\left(\begin{array}{c}m+1\\2\end{array}\right)\left(\begin{array}{c}n+1\\2\end{array}\right)-1.$$
  We can use induction to generalize Proposition 1 to multi-partite case.

\medskip
  {\textbf{Theorem 1}.} Let A be an Hermitian matrix on space $H_{1}\otimes H_{2}\otimes H_{3}\otimes ...\otimes H_{n}$.
  A has tensor production decomposition like $A=\sum_{i=1}^{r}B_{i}^{1}\otimes B_{i}^{2}\otimes ...\otimes B_{i}^{n}$,
  where $B_{i}^{1}, B_{i}^{2}, ..., B_{i}^{n}$ are Hermitian matrices on $H_{1}$, $H_{2}$, ..., $H_{n}$ respectively.

 \medskip
 We now present a practical method to decompose Hermitian matrices into tensor product of Hermitian matrices,
 thus giving a new constructive proof for Theorem 1. Let $E_{ij}^n$ be
 the unit square matrices of size $n\times n$. If it is clear form the context, we will omit the superscript.
To decompose the unit matrix $E_{ij}^{mn}$, we write its indices $i,
j$ uniquely as follows:
$$i=(k-1)n+i', \qquad j=(l-1)n+j', \eqno (6) $$
where $1\leq k, l\leq m$ and $1\leq i', j' \leq n$. Then we have
$$ E_{ij}^{mn}=E_{kl}^m\otimes E_{i'j'}^n.  \eqno (7)$$

Equivalently we can picture the above decomposition as follows. We
first view $E_{ij}^{mn}$ as an $m\times m$ block matrix with each
entry as an $n\times n$ matrix. The resulted block matrix is still a
unit-like matrix where all entries are zero except $(k, l)$-entry,
which is an $n\times n$ unity matrix itself, say $E_{i'j'}$. Then we
immediately have $ E_{ij}^{mn}=E_{kl}^m\otimes E_{i'j'}^n$.

\medskip
{\textbf{Example 1}.} Let $(1+7b)^{-1}\rho_b$ be the density
operator on ${\mathbb C}^2\otimes {\mathbb C}^4$ as follows.
$$
\rho_{b}=\left( \begin{array}{llllllll}
                        b~~0~~0~~0~~~~~~0~~~~b~~0~~~~0 \\
                        0~~b~~0~~0~~~~~~0~~~~0~~b~~~~0 \\
                        0~~0~~b~~0~~~~~~0~~~~0~~0~~~~b\\
                        0~~0~~0~~b~~~~~~0~~~~0~~0~~~~0\\
                        0~~0~~0~~0~~~\frac{1+b}{2}~~~~0~~0~~~\frac{\sqrt{1-b^{2}}}{2}\\
                        b~~0~~0~~0~~~~~~0~~~~b~~0~~~~0\\
                        0~~b~~0~~0~~~~~~0~~~~0~~b~~~~0\\
                        0~~0~~b~~0~~\frac{\sqrt{1-b^{2}}}{2}~~0~~0~~\frac{1+b}{2}\\
                          \end{array}
                          \right). $$
We can decompose $\rho_b$ by the above scheme.
\begin{eqnarray*}
\rho_{b}&=& b\Big(E_{11}^{8}+E_{16}^{8}+E_{22}^{8}+E_{27}^{8}+E_{33}^{8}+E_{38}^{8}+E_{44}^{8}+E_{61}^{8}+E_{66}^{8}+E_{72}^{8}+E_{77}^{8}+E_{83}^{8}\Big)\\
&+&\frac{1+b}{2}(E_{55}^{8}+E_{88}^{8})+\frac{\sqrt{1-b^{2}}}{2}(E_{58}^{8}+E_{85}^{8})\\
&=&b(E_{11}^{2}\otimes E_{11}^{4}+E_{12}^{2}\otimes
E_{12}^{4}+E_{11}^{2}\otimes E_{22}^{4}+E_{12}^{2}\otimes
E_{23}^{4}+E_{11}^{2}\otimes E_{33}^{4}+E_{12}^{2}\otimes
E_{34}^{4}\\
&+&E_{11}^{2}\otimes E_{44}^{4}+E_{21}^{2}\otimes E_{21}^{4}
+E_{22}^{2}\otimes E_{22}^{4}+E_{21}^{2}\otimes
E_{32}^{4}+E_{22}^{2}\otimes E_{33}^{4}+E_{21}^{2}\otimes
E_{43}^{4})\\
&+&\frac{1+b}{2}(E_{22}^{2}\otimes E_{11}^{4}+E_{22}^{2}\otimes
E_{44}^{4})+\frac{\sqrt{1-b^{2}}}{2}(E_{22}^{2}\otimes
E_{14}^{4}+E_{22}^{2}\otimes E_{41}^{4}).
\end{eqnarray*}
For a different decomposition using the singular value
decomposition, the reader is referred to [18].
\medskip

 This decomposition method can be generalized to
Hermitian operators. Let A be a Hermitian matrix, then one can
decompose $A$ into a sum of real and imaginary parts:
$A=B+\sqrt{-1}C$, where $B$ (or $C$) is a symmetric (or
antisymmetric) matrix. Let $\{E_{ij}+E_{ji}\}$ be the basis for the
symmetric matrices, and $\{E_{ij}-E_{ji}\}$ be the basis for the
antisymmetric matrices. It is enough to decompose the basis elements
as tensor products of Hermitian matrices. Roughly speaking, one
writes each basis element $E_{ij}\pm E_{ji}$ of size $mn\times mn$
as a block matrix, then transform it into a tensor product according
to the position where the $1$ or $-1$ appears. The main point is
that we have to consider all Hermitian matrices to factor the basis
elements (cf. the remark after Proposition 1).

Specifically, by modulo $n$ we write the indices $i, j$ uniquely as
in Eq. (6): $i\equiv i' (mod\, n), j\equiv j' (mod\, n)$ and
$k=[(i-1)/n]+1, l=[(j-1)/n]+1 $. Here the representatives for
${\mathbb Z}_n$ are taken to be $\{1, 2, \cdots, n\}$. Then we have
the decomposition
$$ E_{ij}^{mn}+E_{ji}^{mn}=\frac12[(E_{kl}^m+E_{lk}^m)
\otimes (E_{i'j'}^n+E_{j'i'}^n)-\sqrt{-1}(E_{kl}^m-E_{lk}^m) \otimes
\sqrt{-1}(E_{i'j'}^n-E_{j'i'}^n)], \eqno (8)$$

$$ \sqrt{-1}(E_{ij}^{mn}-E_{ji}^{mn})=\frac12[(E_{kl}^m+E_{lk}^m)
\otimes
\sqrt{-1}(E_{i'j'}^n-E_{j'i'}^n)+\sqrt{-1}(E_{kl}^m-E_{lk}^m)
\otimes (E_{i'j'}^n+E_{j'i'}^n)].  \eqno (9)$$

Equivalently we can picture the above decomposition as follows. We
first view $E_{ij}^{mn}\pm E_{ji}^{mn}$ as an $m\times m$ block
matrix $(P_{st})$, where $P_{st}=0$ except $(s, t)=(k, l)$ or $(l,
k)$, and $P_{kl}=P_{lk}^T=E_{i'j'}$. Then we have $
E_{ij}^{mn}+E_{ji}^{mn}=E_{kl}^m\otimes E_{i'j'}^n+ E_{lk}^m\otimes
E_{j'i'}^n$. A simple computation will show that it is also given by
Eq.(8).

\medskip
\textbf{Example 2.} For $f\in[0, 1]$ consider the Werner state [19]
$$\rho=\left(\begin{array}{llll}
                      \frac{1-f}3 & & & \\
                     & \frac{1+2f}6 & \frac{1-4f}6 &  \\
                     & \frac{1-4f}6 & \frac{1+2f}6 &  \\
                  & & &   \frac{1-f}3  \\
\end{array}
\right).  \eqno (10) $$
Then
\begin{eqnarray*}
\rho&=&\frac{1-f}{3}E_{11}^{4}+\frac{1+2f}{6}(E_{22}^{4}+E_{33}^{4})
+\frac{1-4f}6(E_{23}^4+E_{32}^4)+\frac{1-f}{3}E_{44}^{4}\\
&=&\frac{1-f}{3}E_{11}\otimes E_{11}+ \frac{1+2f}{6}(E_{11}\otimes
E_{22}+E_{22}\otimes E_{11}) \\
&+&\frac{1-4f}{12}[(E_{12}+E_{21})\otimes
(E_{21}+E_{12})-i(E_{12}-E_{21})\otimes i(E_{21}-E_{12})]+\frac{1-f}{3}E_{22}\otimes E_{22}.\\
\end{eqnarray*}

\medskip
\textbf{Example 3~} For non-negative a, b, c consider the following
positive semi-definite matrix
$$
\rho=\left( \begin{array}{llllllll}
                        1& 0 & 0 & 0 & 0 & 0 & 0 & 1 \\
                        0 & a& 0 & 0 & 0 & 0 & 0 & 0 \\
                        0 & 0 & b& 0 & 0 & 0 & 0 & 0 \\
                        0 & 0 & 0 & c & 0 & 0 & 0 & 0\\
                        0 & 0 & 0 & 0 & \frac{1}{a} & 0 & 0 & 0\\
                        0 & 0 & 0 & 0 & 0 & \frac{1}{b} & 0 & 0\\
                        0 & 0 & 0 & 0 & 0 & 0 & \frac{1}{c} & 0\\
                        1 & 0 & 0 & 0 & 0 & 0 & 0 & 1\\
                          \end{array}
                          \right).  \eqno (11)
$$
Then we have
\begin{eqnarray*}
\rho&=&E_{11}\ot E_{11}\ot E_{11}+E_{22}\ot E_{22}\ot E_{22}\\
&&+\frac 14S_{12}\ot (S_{12}\ot S_{12}-iA_{12}\ot iA_{12})-\frac 14
iA_{12}\ot (S_{12}\ot iA_{12}-iA_{12}\ot S_{12})\\
&&+aE_{11}\ot E_{11}\ot E_{22}+bE_{11}\ot E_{22}\ot E_{11}+cE_{11}\ot E_{22}\ot E_{22}\\
&&+\frac 1aE_{22}\ot E_{11}\ot E_{11}+\frac 1bE_{22}\ot E_{11}\ot
E_{22}+\frac 1cE_{22}\ot E_{22}\ot E_{11},
\end{eqnarray*}
where $S_{ij}=E_{ij}+E_{ji}$ and $A_{ij}=E_{ij}-E_{ji}$.


\section{Separability of multipartite states}
As we note in the previous section that any Hermitian operator $A$  on a tensor product
 space can be decomposed into a sum of tensor products of Hermitian operators:
 $A=\sum_{i=1}^{r}B_{i}^{1}\otimes B_{i}^{2}\otimes
...\otimes B_{i}^{n}$.
 However the factors $B_{i}^{j}$ are generally not density matrices on $H_{j}$ as they may not be positive operators.
 To answer the question of separability of $A$ one needs to study
 when each factor is non-negative.

 Let $m(A)$ and $M(A)$ denote the smallest and the largest eigenvalues of a Hermitian matrix A. We can transform
 the decomposition into another one so that the smallest
 eigenvalues are nonnegative as follows:
\begin{eqnarray*}
A&=&\sum_{i=1}^{r}B_{i}^{1}\otimes B_{i}^{2}\otimes ...\otimes B_{i}^{n}\\
&=&\sum_{i=1}^{r}\Big(B_{i}^{1}-m(B_{i}^{1})Id_{1}+m(B_{i}^{1})Id_{1}\Big)\otimes \cdots\otimes \Big(B_{i}^{n}-m(B_{i}^{n})Id_{n}+m(B_{i}^{n})Id_{n}\Big)\\
&=&\sum_{i=1}^{t}B_{i}^{'1}\otimes B_{i}^{'2}\otimes ...\otimes
B_{i}^{'n} +q(A)Id_{1}\otimes Id_{2}\otimes ...\otimes Id_{n},
~~~~~~~~~~~~~~~~~~~~~~~~~~~~~~(12)
\end{eqnarray*}
where $B_{i}^{'j}$ are positive semi-definite Hermitian matrices on
$H_{j}$, and each summand has at least one $m(B_{k}^{'j})=0$ but not all (i.e. at least one factor
is the identity $Id_{l}$ on $H_{l}$).

Note that $q(A)$ depends on the decomposition. We
define the \textsl{separability indicator} of A, $S(A)=max(q(A))$ to
be the maximum value of $q(A)$ among all possible decompositions
such as (12). The following result is quoted from [18].
\medskip

{\textbf{Proposition 2}.} Let $A=\sum_i^r B_i\ot C_i$ be a density matrix
on space $H_{1}\otimes H_{2}$. Then $A$ is separable iff the
separability indicator $S(A)\geq 0$. Moreover $S(A)$ satisfies the
following relation $S(A)\leq m(A)$.

\medskip
{ \textbf{Theorem 2}.} Let $A=\sum_{i=1}^{r}B_{i}\otimes C_{i}$ be a
Hermitian operator on the space $H_1\ot H_2$, then $q(A)$ is given
by
\begin{eqnarray*}
q(A)&=&\sum_{m(B_i)\geq 0, m(C_i)\geq 0}m(B_{i})m(C_{i})+\sum_{m(B_i)<0}m(B_{i})M(C_{i})\\
&+&\sum_{m(C_i)<0}M(B_{i})m(C_{i})-\sum_{m(B_i)<0,m(C_i)<0}m(B_{i})m(C_{i}), \hskip 1.5in\hfill(13)
\end{eqnarray*}
and bounded by
\begin{eqnarray*}
 q(A)&\geq &M(A)-\sum_{i=1}^{r}\Big[\Big(M(B_{i})-m(B_{i})\Big)\Big(M(C_{i})-m(C_{i})\Big)+M\Big(m(C_{i})B_{i}\Big)\\
&-&m\Big(m(C_{i})B_{i}\Big)+M\Big(m(B_{i})C_{i}\Big)-m\Big(m(B_{i})C_{i}\Big)\Big].    \hskip 1.7in(14)
\end{eqnarray*}

{\textbf{Proof.}} For any Hermitian matrix $P$ we define the
 operation $P'$ by shifting with the minimum eigenvalue: $P'=P-m(P)I$. We can
 rewrite Eq. (12)
\begin{eqnarray*}
A&=&\sum_{i=1}^{r}B_i'\otimes C_i'+\Big(m(C_{i})B_i'-m((m(C_{i})B_i')I_{m}\Big)\otimes I_{n}\\
&+&I_{m} \otimes
\Big(m(B_{i})C_i'-m(m(B_{i})C_i')I_{n}\Big)+q(A)I_{m}\otimes
I_{n},
\end{eqnarray*}
where
$$q(A)=\sum_i\left(m\Big(m(C_{i})B_i'\Big)+m\Big(m(B_{i})C_i'\Big)+m(B_{i})m(C_{i})\right).$$

We observe that for any real $s$
$$m(sP)=\frac{s+|s|}2m(P)+\frac{s-|s|}2M(P),\quad M(sP)=\frac{s+|s|}2M(P)+\frac{s-|s|}2m(P). \eqno(15)$$
It then follows that
\begin{eqnarray*}
&&q(A)=\sum_i\left(m\Big(m(C_{i})B_i'\Big)+m\Big(m(B_{i})C_i'\Big)+m(B_{i})m(C_{i})\right)\\
&=&\sum_{m(B_i)\geq 0, m(C_i)\geq 0}m(B_{i})m(C_{i})+\sum_{m(B_i)<0, m(C_i)\geq 0}m(B_{i})M(C_{i})+\sum_{m(B_i)\geq 0,
m(C_i)<0}M(B_{i})m(C_{i})\\
&&+\sum_{m(B_i)<0,m(C_i)<0}\Big(m(C_{i})M(B_{i})+m(B_{i})M(C_{i})-m(B_{i})m(C_{i})\Big)\\
&=&\sum_{m(B_i)\geq 0, m(C_i)\geq 0}m(B_{i})m(C_{i})+\sum_{m(B_i)<0}m(B_{i})M(C_{i})\\
&&+\sum_{m(C_i)<0}M(B_{i})m(C_{i})-\sum_{m(B_i)<0,m(C_i)<0}m(B_{i})m(C_{i}).
\end{eqnarray*}

Now we notice that for any matrix $P$ and any real number $r$,
$M(P-rI)=M(P)-r$, $m(P-rI)=m(P)-r$, from which it follows that
\begin{eqnarray*}
&&M(B_i')=M(B_{i}-m(B_{i})I_{m})=M(B_{i})-m(B_{i}),\\
&&M\Big(m(C_{i})B_i'-m(m(C_{i})B_i')I_{m}\Big)
=M\Big(m(C_{i})B_{i}\Big)
-m\Big(m(C_{i})B_{i}\Big). \hskip 1.5in (16)
\end{eqnarray*}

On the other hand it is well-known that $M(A+B)\leq M(A)+M(B)$ (see
[20]).  Thus taking the maximum eigenvalues, we get
\begin{eqnarray*}
q(A)&\geq &M(A)-\sum_{i=1}^{r}\Big[\Big(M(B_{i})-m(B_{i})\Big)\Big(M(C_{i})-m(C_{i})\Big)+M\Big(m(C_{i})B_{i}\Big)\\
&-&m\Big(m(C_{i})B_{i}\Big)+M\Big(m(B_{i})C_{i}\Big)-m\Big(m(B_{i})C_{i}\Big)\Big],
\end{eqnarray*}
 which completes the proof.

In the last part of the proof if we take minimum eigenvalues we will get the known inequality
 $m(A)\geq q(A)$ (using $m(A+B)\geq m(A)+m(B)$).

 We remark that the above lower bound is different from that in
 [18]. To better understand our lower bounds, we consider the
 special case when all
 factors are non-negative matrices, then $m(m(B_i)C_i)=m(B_i)m(C_i)$ etc. Then
 it follows that
$$
q(A)\geq M(A)-\sum_{i=1}^{r}\Big[M(B_{i})M(C_{i})-m(B_{i})m(C_{i})\Big]. \eqno(17)
$$

While the other extreme case is when all factors are negative, then
$$
q(A)\geq
M(A)-\sum_{i=1}^{r}\Big[\Big(M(B_{i})-2m(B_{i})\Big)\Big(M(C_{i})-2m(C_{i})\Big)-m(B_{i})m(C_{i})\Big].
$$

When the factors $B_i$ or $C_i$ are not all non-negative, we have
\begin{eqnarray*}
q(A)&\geq& M(A)-\sum_{m(B_i)\geq 0, m(C_i)\geq 0}\Big[M(B_{i})M(C_{i})-m(B_{i})m(C_{i})\Big]\\
&-&\sum_{m(B_i)<0, m(C_i)\geq 0}\Big[\Big(M(B_{i})-2m(B_i)\Big)M(C_{i})+m(B_{i})m(C_{i})\Big]\\
&-&\sum_{m(B_i)\geq 0, m(C_i)<0}\Big[M(B_{i})\Big(M(C_{i})-2m(C_i)\Big)+m(B_{i})m(C_{i})\Big]\\
&-&\sum_{m(B_i)<0,
m(C_i)<0}\Big[\Big(M(B_{i})-2m(B_{i})\Big)\Big(M(C_{i})-2m(C_{i})\Big)-m(B_{i})m(C_{i})\Big].~~~~~~~~~(18)
\end{eqnarray*}

The above result can be generalized to multipartite states.

\vskip 0.2in { \textbf{Theorem 3}.} Let $A=\sum_i^r B_i\ot C_i\ot D_i$ be a density matrix
on space $H_{1}\otimes H_{2}\otimes H_{3}$. Then $A$ is separable if
and only if the separability indicator $S(A)\geq 0$. Moreover $S(A)$
satisfies the following relations
$$S(A)\leq m(A).\eqno(19)$$


\begin{eqnarray*}
q(A)&\geq&M(A)-\sum_{i=1}^{r}\Big[\Big(M(B_{i})-m(B_{i})\Big)\Big(M(C_{i})
-m(C_{i})\Big)\Big(M(D_{i})-m(D_{i})\Big)\\
&+&M\Big(m(B_{i})m(D_{i})C_{i}\Big)-m\Big(m(B_{i})m(D_{i})C_{i}\Big)+M\Big(m(C_{i})m(D_{i})B_{i}\Big)\\
&-&m\Big(m(C_{i})m(D_{i})B_{i}\Big)+M\Big(m(B_{i})m(C_{i})D_{i}\Big)-m\Big(m(B_{i})m(C_{i})D_{i}\Big)\\
&+&M\Big(m(m(B_{i})C_{i})D_{i}-m(B_{i})m(C_{i})D_{i}\Big)-m\Big(m(m(B_{i})C_{i})D_{i}-m(B_{i})m(C_{i})D_{i}\Big)\\
&+&M\Big(m(m(C_{i})B_{i})D_{i}-m(C_{i})m(B_{i})D_{i}\Big)-m\Big(m(m(C_{i})B_{i})D_{i}-m(C_{i})m(B_{i})D_{i}\Big)\\
&+&M\Big(m(m(D_{i})B_{i})C_{i}-m(D_{i})m(B_{i})C_{i}\Big)-m\Big(m(m(D_{i})B_{i})C_{i}-m(D_{i})m(B_{i})C_{i}\Big)\\
&+&\Big(M(m(D_{i})B_{i})
-m(m(D_{i})B_{i})\Big)\Big(M(C_{i})-m(C_{i})\Big)\\
&+&\Big(M(m(C_{i})B_{i})-m(m(C_{i})B_{i})\Big)\Big(M(D_{i})-m(D_{i})\Big)\\
&+&\Big(M(m(B_{i})C_{i})-m(m(B_{i})C_{i})\Big)\Big(M(D_{i})-m(D_{i})\Big)\Big].
\hskip 1.5in (20)
\end{eqnarray*}
The idea of the proof will be similar to that of Theorem 2 and is
included in the Appendix. More generally we can use the same idea to
give similar results for multi-partite cases.
\medskip

{\textbf{Theorem 4}.} Let
$A$ be a $k$-partite mixed state on space $H_{1}\otimes
H_{2}\otimes ...\otimes H_{k}$, then $A$ has a tensor decomposition
into Hermitian operators in the form
$A=\sum_{i=1}^{r}B_{i}^{1}\otimes B_{i}^{2}\otimes ...\otimes
B_{i}^{k}$ and is separable if and only if the separability indicator
$S(A)\geq 0$. Moreover $S(A)$ satisfies the following relation
$$S(A)\leq m(A).\eqno(21)$$
When all factors are non-negative, we have
\begin{eqnarray*}
q(A)&=&\sum_i^r m(B_{i}^{1})m(B_{i}^{2})\cdots
m(B_{i}^{k})\\
&\geq &  M(A)-\sum _{i=1}^{r}\Big[M(B_{i}^{1})M(B_{i}^{2})\cdots
M(B_{i}^{k})-m(B_{i}^{1})m(B_{i}^{2})\cdots
m(B_{i}^{k})\Big]. \hskip 1in (22)
\end{eqnarray*}

\section{Conclusion}

We have developed a criterion to judge whether a multi-partite
density operator is separable. Our idea is first to decompose the
density operator into a sum of tensor product of Hermitian
operators. We give a new and practical way to decompose any
Hermitian operator into tensor product of Hermitian operators in
multi-partite cases. Unlike the numerical method [17] and the method
of singular value decomposition [18] our new method is completely
elementary and algebraic. Using the decomposition we can rewrite it into a
tensor product of positive operators plus a scaler operator, which
is called the separability indicator. The separability indicator
provides a new mechanism to measure the quantum entanglement of the
density operator. We derive some bound to estimate the scalar or
separability indicator. Our inequalities are expressed in terms of
eigenvalues of the summands, and in some case they are sufficient to
tell if the separability indicator is non-negative, thus shows that
the density operator is separable. As our method relies on how the
operator is decomposed, it is usually difficult to compute the
separability indicator exactly. We hope our estimates will shed more
light on the separability problem.

\bigskip
\centerline{\bf Acknowledgments}

We are grateful to the referees' stimulating comments which lead to clarification and
simplification of the arguments in the paper. Jing thanks the support of NSA grant H98230-06-1-0083 and
NSFC's Overseas Distinguished Youth Grant.

\bigskip

\bibliographystyle{amsalpha}

\section{Appendix}

{\textbf{Proof of Theorem 3.}} The idea of the proof is similar to
that of Theorem 2. Recall the meaning of operation
$P'=P-m(P)I$, and we have
\begin{eqnarray*}
A&=&\sum_{i=1}^{r}B_{i}\otimes C_{i}\otimes D_{i}=\sum_{i=1}^{r}B_i'\otimes C_i'\otimes D_i'+\sum_{i=1}^{r}\Big(m(D_{i})B_i'-m  (m(D_{i})B_i')I_{m} \Big)\otimes C_i'\otimes I_{k}\\
&+&\sum_{i=1}^{r}I_{m}\otimes \Big(m(m(D_{i})B_i')C_i'-m(m(m(D_{i})B_i')C_i')I_{n} \Big) \otimes I_{k}\\
&+&\sum_{i=1}^{r}I_{m}\otimes \Big[m(B_{i})m(D_{i})C_i'-m\Big(m(B_{i})m(D_{i})C_i'\Big)I_{n} \Big]\otimes I_{k}\\
&+&\sum_{i=1}^{r}I_{m}\otimes \Big(m(B_{i})C_i'-m(m(B_{i})C_i')I_{n}\Big)\otimes D_i'\\
&+&\sum_{i=1}^{r}\Big(m(C_{i})B_i'-m(m(C_{i})B_i')I_{m}\Big)\otimes I_{n}\otimes D_i'\\
&+&\sum_{i=1}^{r}I_{m}\otimes I_{n}\otimes \Big(m(m(B_{i})C_i')D_i'-m(m(m(B_{i})C_i')D_i')I_{k}\Big)\\
&+&\sum_{i=1}^{r}\Big[m(C_{i})m(D_{i})B_i'-m\Big(m(C_{i})m(D_{i})B_i'\Big)I_{m}\Big]\otimes I_{n}\otimes I_{k}\\
&+&\sum_{i=1}^{r}I_{m}\otimes I_{n}\otimes \Big(m(m(C_{i})B_i')D_i'-m(m(m(C_{i})B_i')D_i')I_{k}\Big)\\
&+&\sum_{i=1}^{r}I_{m}\otimes I_{n}\otimes
\Big[m(B_{i})m(C_{i})D_i'-m\Big(m(B_{i})m(C_{i})D_i'\Big)I_{k}\Big]+q(A)I_{m}\otimes
I_{n}\otimes I_{k},~~~~(23)
\end{eqnarray*}
where
\begin{eqnarray*}
q(A)&=&\sum_{i=1}^{r}m\Big(m(m(D_{i})B_i')C_i'\Big)+\sum_{i=1}^{r}m\Big(m(m(B_{i})C_i')D_i'\Big)+\sum_{i=1}^{r}m\Big(m(m(C_{i})B_i')D_i'\Big)\\
&+&\sum_{i=1}^{r}m\Big(m(C_{i})m(D_{i})B_i'\Big)+\sum_{i=1}^{r}m\Big(m(B_{i})m(D_{i})C_i'\Big)+\sum_{i=1}^{r}m\Big(m(B_{i})m(C_{i})D_i'\Big)\\
&+&\sum_{i=1}^{r}m(B_{i})m(C_{i})m(D_{i})\\
&=&\sum_{i=1}^{r}m\Big(m(m(D_{i})B_{i})C_{i}-m(D_i)m(B_i)C_{i}\Big)+\sum_{i=1}^{r}m\Big(m(m(B_{i})C_{i})D_{i}-m(B_i)m(C_i)D_{i}\Big)\\
&+&\sum_{i=1}^{r}m\Big(m(m(C_{i})B_{i})D_{i}-m(B_i)m(C_i)D_{i}\Big)-\sum_{i=1}^{r}m\Big(m(D_{i})B_{i}\Big)m(C_{i})\\
&-&\sum_{i=1}^{r}m\Big(m(B_{i})C_{i}\Big)m(D_{i})-\sum_{i=1}^{r}m\Big(m(C_{i})B_{i}\Big)m(D_{i})+\sum_{i=1}^{r}m(B_{i})m(C_{i})m(D_{i}),~~~~(24)
\end{eqnarray*}
where we have used similar identities like Eq. (16). Now we would like
to consider eight possible signs of $m(B_i), m(C_i), m(D_i)$, and we
use $+, -, +$ to denote the
subset $\{i|m(B_i)\geq 0, m(C_i)<0, m(D_i)\geq 0\}$ etc.
to simplify the notation. By Eq. (15) it
follows that
\begin{eqnarray*}
q(A)&=&\sum_{+,+,+}m(B_i)m(C_i)m(D_i)\\
&+&\sum_{-,+,+}m(B_i)\Big[M(C_i)M(D_i)-M(C_i)m(D_i)-m(C_i)M(D_i)-m(C_i)m(D_i)\Big]\\
&+&\sum_{+,-,+}m(C_i)\Big[M(B_i)M(D_i)-m(B_i)M(D_i)-M(B_i)m(D_i)-m(B_i)m(D_i)\Big]\\
&+&\sum_{+,+,-}m(D_i)\Big[M(B_i)M(C_i)-m(B_i)M(C_i)-M(B_i)m(C_i)-m(B_i)m(C_i)\Big]\\
&+&\sum_{-,-,+}\Big[m(B_i)\Big(M(C_i)M(D_i)-M(C_i)m(D_i)-m(C_i)M(D_i)\Big)\\
&+&m(C_i)\Big(M(B_i)M(D_i)-m(B_i)M(D_i)-M(B_i)m(D_i)\Big)\Big]\\
&+& \sum_{-,+,-}\Big[m(B_i)\Big(M(C_i)M(D_i)-M(C_i)m(D_i)-m(C_i)M(D_i)\Big)\\
&+&m(D_i)\Big(M(B_i)M(C_i)-m(B_i)M(C_i)-M(B_i)m(C_i)\Big)\Big]\\
&+& \sum_{+, -, -}\Big[m(C_i)\Big(M(B_i)M(D_i)-m(B_i)M(D_i)-M(B_i)m(D_i)\Big)\\
&+&m(D_i)\Big(M(B_i)M(C_i)-m(B_i)M(C_i)-M(B_i)m(C_i)\Big)\Big]\\
&+&\sum_{-,-,-}\Big(m(D_i)M(C_i)M(B_i)+m(B_i)M(C_i)M(D_i)
+m(C_i)M(D_i)M(B_i)\\
&-&2m(B_i)m(C_i)M(D_i)-2m(B_i)m(D_i)M(C_i)-2m(C_i)m(D_i)M(B_i)\\
&& \qquad +m(B_i)
m(C_i)m(D_i)\Big).
\end{eqnarray*}

The above expression leads to an easy proof of the criterion: if $A$
is separable, then all the factors are non-negative and $S(A)\geq
q(A)=\sum_{i}m(B_i)m(C_i)m(D_i)\geq 0.$ The converse is immediate.

If we take minimum eigenvalues to the decomposition (12), we will
get
$$m(A)\geq \sum_im(B_{i}^{'1})m(B_{i}^{'2})\cdots m(B_{i}^{'n}) +q(A)=q(A). \eqno(25)
$$

Next using similar identities as Eq. (16) we get identities like
\vskip 0.1in $M\Big(m(D_{i})B_i'-m (m(D_{i})B_i')I_{m}
)\Big)=M\Big(m(D_{i})B_{i}\Big)-m\Big(m(D_{i})B_{i}\Big) $,

\begin{eqnarray*}
M\Big(m(m(D_{i})B_i')C_i'-m(m(m(D_{i})B_i')C_i')I_{n}\Big)
&=&M\Big(m(m(D_{i})B_{i})C_{i}-m(D_{i})m(B_{i})C_{i}\Big)\\
&-&m\Big(m(m(D_{i})B_{i})C_{i}-m(D_{i})m(B_{i})C_{i}\Big).
\end{eqnarray*}

Thus we have

\begin{eqnarray*}
M(A)&\leq&\sum_{i=1}^{r}\Big[\Big(M(B_{i})-m(B_{i})\Big)\Big(M(C_{i})
-m(C_{i})\Big)\Big(M(D_{i})-m(D_{i})\Big)\\
&+&M\Big(m(B_{i})m(D_{i})C_{i}\Big)-m\Big(m(B_{i})m(D_{i})C_{i}\Big)+M\Big(m(C_{i})m(D_{i})B_{i}\Big)\\
&-&m\Big(m(C_{i})m(D_{i})B_{i}\Big)+M\Big(m(B_{i})m(C_{i})D_{i}\Big)-m\Big(m(B_{i})m(C_{i})D_{i}\Big)\\
&+&M\Big(m(m(B_{i})C_{i})D_{i}-m(B_{i})m(C_{i})D_{i}\Big)-m\Big(m(m(B_{i})C_{i})D_{i}-m(B_{i})m(C_{i})D_{i}\Big)\\
&+&M\Big(m(m(C_{i})B_{i})D_{i}-m(C_{i})m(B_{i})D_{i}\Big)-m\Big(m(m(C_{i})B_{i})D_{i}-m(C_{i})m(B_{i})D_{i}\Big)\\
&+&M\Big(m(m(D_{i})B_{i})C_{i}-m(D_{i})m(B_{i})C_{i}\Big)-m\Big(m(m(D_{i})B_{i})C_{i}-m(D_{i})m(B_{i})C_{i}\Big)\\
&+&\Big(M(m(D_{i})B_{i})
-m(m(D_{i})B_{i})\Big)\Big(M(C_{i})-m(C_{i})\Big)\\
&+&\Big(M(m(C_{i})B_{i})-m(m(C_{i})B_{i})\Big)\Big(M(D_{i})-m(D_{i})\Big)\\
&+&\Big(M(m(B_{i})C_{i})-m(m(B_{i})C_{i})\Big)\Big(M(D_{i})-m(D_{i})\Big)\Big]+q(A),~~~~~~~~~~~(26)
\end{eqnarray*}
which completes the proof of Theorem 3.

We remark that when all
 factors are non-negative matrices, then
 it follows that
$$
q(A)\geq
M(A)-\sum_{i=1}^{r}\Big[M(B_{i})M(C_{i})M(D_{i})-m(B_{i})m(C_{i})m(D_{i})\Big].
\eqno(27)
$$

While the other extreme case is when all factors are negative, then
$$
q(A)\geq
M(A)-\sum_{i=1}^{r}\Big[\Big(M(B_{i})-3m(B_{i})\Big)\Big(M(C_{i})-3m(C_{i})\Big)\Big(M(D_{i})-3m(D_{i})\Big)-m(B_{i})m(C_{i})m(D_{i})\Big].
$$

{\textbf{Proof of Theorem 4.}} The proof is by an easy induction as those of Theorems 2 and 3.
Some details are already offered in Eqs (25) and (26).

\end{document}